\documentclass[runningheads]{llncs}
\usepackage{graphicx}
\usepackage{xcolor}
\usepackage{textcomp, color, soul}

\definecolor{editCol}{rgb}{0.0, 0.0, 255.0}

\begin{document}
\title{A User Study to Evaluate a Web-based Prototype for Smart Home Internet of Things Device Management}
%
%
\author{Leena Alghamdi \inst{1} \and Ashwaq Alsoubai\inst{1} \and Mamtaj Akter\inst{1} \and
Faisal Alghamdi \inst{1} \and
Pamela Wisniewski\inst{2}
}
\authorrunning{L. Alghamdi et al.}
%
\institute{University of Central Florida, Orlando FL 32816, USA 
\\
\email{[leenaalghamdi, atalsoubai, mamtaj.akter, faisal.ramzi]@knights.ucf.edu}\\
\email{pamwis@ucf.edu}
}
\maketitle              
\begin{abstract}

 With the growing advances in the Internet of Things (IoT) technology,  IoT device management platforms are becoming increasingly important. We conducted a web-based survey and usability study with 43 participants who use IoT devices frequently to: 1) examine their smart home IoT usage patterns and privacy preferences, and 2) evaluate a web-based prototype for smart home IoT device management. We found that participants perceived privacy as more important than the convenience afforded by the IoT devices. Based on their average scores of the privacy vs. convenience importance, participants with low privacy and low convenience significantly reported less privacy control and convenience preferences than participants with high privacy and high convenience. Overall, all participants were satisfied about the proposed website prototype and their actual usability evaluation demonstrated good understanding of the website features.  
 This paper provides an empirical examination of  the privacy versus convenience trade-offs smart home users make when managing their IoT devices.

\end{abstract}

\section{Introduction}

According to MediaPost \cite{martin_smart_nodate}, 69\% of households in the United States own at least one smart Internet of Things (IoT) device, while 12\% of those (about 22 million homes) own multiple devices. Despite the widespread proliferation of IoT smart home technologies, there are several concerns around the data privacy and management of these IoT devices. For example, people do not feel comfortable with third-parties using their sensitive data \cite{naeini_privacy_2017}.  However, people use third-party platforms to manage their smart IoT devices,  even though these platforms invade users’ surroundings and capture their sensitive information without permission \cite{balliu_securing_2019}. The reasons behind this is that people may not understand the extent of the data collection by the third party, or people may think the trade-off is worth it for the added convenience \cite{molla_people_2019}. In other cases, some people may not care enough about their personal privacy to be concerned about data leakages \cite{molla_people_2019}. Thus, it is important to further understand the trade-off between privacy and convenience in the context of smart home IoT device management.

There are various platforms that already exist that aim to provide a centralized management platform for IoT devices. For example, Silva et al. \cite {silva_m4dniot-networks_2019}, proposed a system management tool for devices and networks in IoT with user interface (M4DN.IoT), and this system provides information about connected devices and networks. It supports both automatic IoT networks management and user interface. Although the existing systems provide a management platform for IoT devices, they come with certain limitations. For example, some of the platforms are proposed for smart devices controlling purposes only, they do not preserve the users privacy and provide no mechanisms for protecting sensitive information. At the same time, though, it remains unclear whether a privacy-focused IoT device management platform is actually improving IoT users' privacy perceptions by sufficiently helping them to manage their smart home IoT devices based on their privacy preferences and convenience of usage. Consequently, we asked the following high level research questions:

\begin{itemize}
  \item \textbf{RQ1:} \textit{Do smart home IoT device users generally value their privacy versus convenience more?}
 \item \textbf{RQ2:} \textit{Based on their preference towards privacy versus convenience, how does this influence their decisions to protect their privacy?} 
\item  \textbf{RQ3:} \textit{Does this preference influence how they evaluate a web-based prototype for centralized IoT smart home management?}
\end{itemize}

To address these research questions, we first developed a web-based prototype as a centralized location for users to manage their IoT smart home devices. This prototype is intended to enable IoT users to gather all of their smart devices into a single platform and effortlessly manage them, while protecting their privacy based on their preferences when managing their devices. We gave participants several tasks to complete using the prototype. We then conducted a web-based survey with 43 adults to evaluate the prototype and answer survey questions about their preferences towards privacy versus convenience, as well as their privacy control, privacy preference, convenience preferences and their website satisfaction.

Overall, we found that the majority of our participants valued both privacy and convenience (RQ1). Most participants (37.2\%, N=16) valued both privacy and convenience, followed by convenience over privacy (27.9\%, N=12), neither privacy nor convenience (23.3\%, N=10), and privacy over convenience (11.6\%, N=5). However, within-subjects, we found significant differences between privacy and convenience importance. Specifically, participants were more concerned about privacy than convenience when using IoT devices.  For RQ2, we found significant differences based on the privacy vs. convenience profiles, such that participants who were in the low privacy/ low convenience group significantly reported less Privacy Control and Convenience Preferences than the group of high privacy/ high convenience. For RQ3,participants generally were satisfied about our website with the website organization, ease of website navigation, and the user-interface when they experience it. Participants' responses towards the website satisfaction also reflected on their website usage. Most of our participants ($N=35, 92.11\%$) could follow our website usage instructions and could navigate to the website pages to perform the activities that we asked them to do. This gave us the further understanding of how our website design appeared easy to learn and use to our participants.

\par This study contributes to the field of IoT smart home management by evaluating users' perception of using one platform to manage IoT devices while protecting their privacy based on their preferences, as IoT device management platform ensure that users' privacy requirements are met. Also, it provides a clear ideas about privacy and convenience preferences of smart home users when using such a platform. This paper is organized as follows: Section 2 reviews the background of smart home IoT privacy issues, then outlines some of the research contributions to overcome these issues, also, it discusses some of the IoT devices management mechanisms. Section 3 describes the process of designing and implementing our proposed prototype. Section 4 describes our methodologies with the details regarding our analysis approach. Section 5  highlights the findings, and section 6 discusses the key findings, outlines the limitations and provides an outlook of the future work that needs to be conducted in this area. Finally, Section 7 concludes the research.

\section{Related Work}

A Pew research ~\cite{noauthor_smart_2016} reported that 55\% of the smart device users find it unacceptable that smart home devices collect their sensitive personal information (e.g., precise location, communication patterns, physical movements, and so on). Although smart devices bring conveniences and monetary benefits for home, many recent work reported that these devices have become cause of concerns to the users ~\cite{lau_alexa_2018,zheng_user_2018,noauthor_living_nodate,mccreary_contextual_2016}. For example, Arabo et al. in \cite{arabo_privacy_2012} studied how smart home network users' online security, safety and privacy can be compromised. They summarized the threats in several categories, i.e., identity theft, social threat, online safety, data security, cyber attacks. Some other research works also reported that users in general are not aware of these privacy threats since they are not informed how and to what extent their personal information is being accessed ~\cite{zheng_user_2018,kroger_personal_2021,saeidi_if_2021,gupta_two-fold_2022}. Additionally, smart home users often share their devices with trusted people who live outside of their home \cite{tabassum_smart_2020}. Thus, researchers have suggested to design IoT management tools that allow users to have a transparency on the data that get collected by their smart home devices ~\cite{lau_alexa_2018,zheng_user_2018,saeidi_if_2021}. To this end, Yan et al. designed a smart home device data monitoring mechanism titled RestThing \cite {zhang_survey_2014} that enabled users to monitor the status of their physical and technological resources including their collected information. Moreover, Delicato et al. \cite {delicato_platform_2014} proposed a web-based paradigm (EcoDiF) that aims to offer a platform that provides users a real-time data monitoring and visualization. Although these proposed solutions combined physical devices with IoT networks and provided web services to users to monitor; some recent work revealed that IoT management tools still require to be more user centric so that users can have the agency over their own privacy management and this self privacy management may benefit them to become more aware and confident about their data privacy management.


Many researchers have proposed solutions to allow IoT users to manage their devices via web pages or web/mobile applications. For example, Piyare in \cite {piyare_bluetooth_2011} proposed a low cost, flexible, and Web-server-based solution to smart home device control. However, their system is only for switching and controlling home appliances and devices through an Android-based app that can work only by using Android Smartphone or Android Tablet, and their system did not help users in preserving their privacy. Similar to this work, \cite {kunal_smart_2016} and \cite {shrestha_web_2017} also designed and developed Android based web apps that allowed users to control the smart home devices through it and interact with devices remotely using Android Smartphones and it provides voice command functionalities, security, and save energy as well. Similar to the Piyare’s system, this system is an Android-based app that works only on Android smartphones. While all these research focused on allowing users to monitor and control the data collected by the smart home devices on specific platform (Android), we identified that there are more research needed to understand how users would be benefited if we could design a web application (that can be accessed from web browsers) for such mechanisms. It is also important to know how users' privacy perception, sense of privacy importance impact on their behavior towards such management tool. Our research makes a contribution to this end by developing a web-based prototype as a centralized platform for users to remotely manage and review the data collected by their IoT smart home devices and evaluating this prototype from participants' perception of privacy concern and convenience of usage. In the next section, we provide a detailed overview of the design of our developed web prototype.

 \begin{figure}[h]
   \includegraphics[width=1\linewidth ]{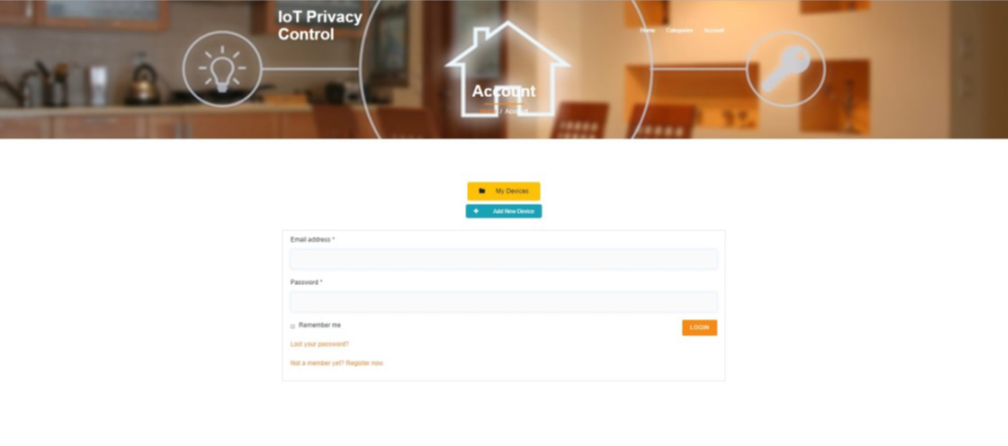}
   \caption{The Login page}
     \label{fig:login}
   \setlength{\belowcaptionskip}{0pt}
   \setlength{\abovecaptionskip}{-30pt}
 \end{figure} 

\section{Design of an IoT Device Management System}

We designed and developed a web-based IoT device management interface that acted as a centralized portal for IoT device management. The website included the following webpages and capabilities:

\subsection{Account Login}
The Account Login (Figure-\ref{fig:login}) page allowed users to log in to their accounts by signing up as a new user with the following required information: Email Address and Password. The purpose of this page was to ensure users that the website was secure and that sensor-based information presented by the website would not be accessible to the public.

\subsection{Device Categories Page}
\begin{figure}[h]
  \includegraphics[width=1\linewidth ]{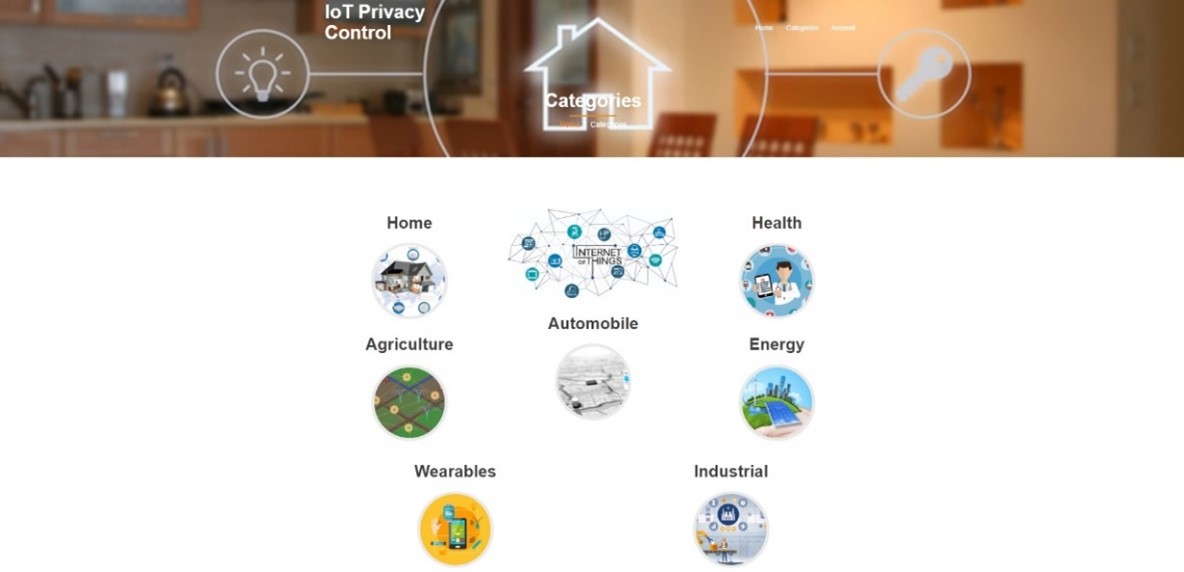}
  \caption{The Category page}
    \label{fig:category}
  \setlength{\belowcaptionskip}{0pt}
  \setlength{\abovecaptionskip}{-30pt}
\end{figure}

Next, the Category Page (Figure-\ref{fig:category}) was organized into seven categories of IoT device types (i.e., home, health, agricultural, automobile, wearable, energy, and industrial). These categories were selected based on several factors, including importance to user’s daily life and their coverage on a large number of IoT devices and sensors. For the purposes of this study, we only implemented a temperature and pressure sensor under the `Home' category. 

\subsection{IoT Device Management Page}
\begin{figure}[h]
  \includegraphics[width=1\linewidth ]{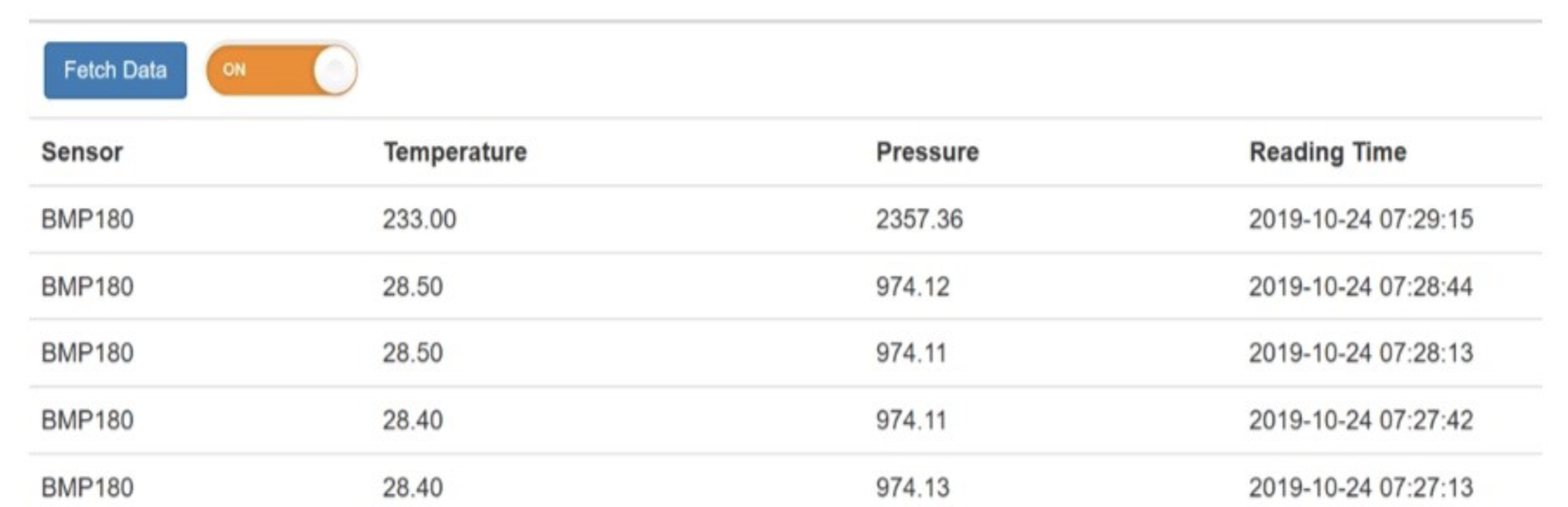}
  \caption{An IoT device (Temperature and Pressure Sensor) Page}
    \label{fig:sensor}
  \setlength{\belowcaptionskip}{0pt}
  \setlength{\abovecaptionskip}{-30pt}
\end{figure}

This prototype allowed users to interact with an IoT device that measured the room temperature and air pressure. On the IoT Device Management Page for the temperature and pressure sensor, users were able to: 1) review the device status and history, and 2) control the device. Our web-based prototype retrieved the temperature readings the database and displayed them in a human-readable format to our participants in the Temperature and Pressure Sensor page under the Home category (Figure - \ref{fig:sensor}). Through this page, users also could turn on or off the device. Lastly, a Home Page allowed users to navigate to all other pages.

\subsection{System Architecture}
We created the web-based interface using WordPress, an open-source program, as a Content Management System (CMS) that includes plugin architecture and template system features. We also implemented a sensor (BMP180) that was connected to an ESP32 controller board and communicated with a mySQL database. The BMP180 was chosen because was low-cost, and it enabled us to measure real-time temperature and pressure, also to estimate the altitude that affects the pressure. Also, the ESP32 was chosen because it was a low-powered system with integrated Wi-Fi, which is universally used for IoT applications. We programmed the ESP32 controller board with Arduino IDE, using a PHP script to insert data into our MySQL database that provides enough storage capacity to store the needed data.The wires that were used for wiring the BMP180 to the ESP32, the I2C pins are GPIO 22: SCL (SCK), and GPIO 21: SDA (SDI). Consequently, all the values from the BMP180 sensor, such as temperature and pressure of that particular area, were shown on the website via connecting to the MySQL database. Our website displayed the BMP180 sensor readings and timestamps from the database to allow data visualization on the website, as shown in (Figure - \ref{fig:sensor}).

\section{Methods}
Below, we provide an overview of our study methodology, the details regarding our analysis approach, and then explain our recruitment strategy.

\subsection{Study Overview}
The primary purpose of our study was to understand participants’ perception of digital privacy versus convenience and to evaluate our prototype keeping these preferences in mind. Therefore, we designed our user study to include two distinct phases: 1) A Web-based survey that included a questionnaire regarding convenience and privacy perceptions for smart device usage, 2) A guided exploration of the prototype with pre-defined tasks.

 \subsection{Study Procedure}
The study started with asking the participants to answer the eligibility screening questions (whether they were at least 18 years old) and sign the statement of informed consent. Participants then provided their demographic information (e.g., age, education level). Next, a web-based survey that consisted of newly developed measures related to their perception of privacy and convenience of smart device usage (Table-\ref{tab:satisfaction} -\ref{tab:privacy-measures} in Appendix A and B). Participants were then asked to perform a specific set of tasks using different pages of our web prototype. Participants were instructed to browse to our web application from their web browsers using any device (e.g., smart phones, tablets, or computers). Since we aimed to evaluate participants' actual usage of the website prototype, the participants were asked to log into the website as a user. Next, we asked the participants to complete the following tasks: 1) Discover how many smart devices are connected to this website, 2) Turn on the temperature and pressure sensor and review the temperature readings, 3) Turn off the temperature and pressure sensor to pause collecting information and review the temperature readings. Once the participants' interactions with the prototype were completed, participants were then asked to complete another set of survey questions regarding their level of satisfaction on the web prototype usability. Participants took between thirty minutes to one and a half hours to complete the study. Next, we describe the survey measured created for this study. 

\subsection{Privacy and Convenience Constructs}
In this study, we developed survey constructs to measure the importance and preferences aspects of privacy and convenience when using IoT devices. All measures were based on a 5-point Likert scale from 1 to 5 (scale 1 = Not important at all, scale 5 = Very Important). All measures reported satisfactory (higher than 0.7) Cronbach's alpha, which measures the internal consistency of survey constructs \cite{cronbach1955construct} as listed in the descriptive statistics (Table \ref{stats}). The following subsections will describe each measure in more detail. 

\subsubsection{Privacy Importance.}
Privacy is one of the most concerning factors that affect users' decisions to  adopt and/or use IoT devices \cite{weinberg2015internet}. Therefore, we were interested to measure the users' perception about their privacy in the context of IoT devices to compare it with their perceptions about the convenience afforded by these devices. Therefore, a Privacy Importance measure was developed to measure users' perception about the importance of protecting their privacy when using smart devices. This measure included one question for the participants to rate how important their privacy when they use IoT devices. 

\subsubsection{Convenience Importance.}
IoT users consider the tradeoffs between privacy and convenience when using smart devices \cite{weinberg2015internet}. Therefore, for this study, we included these two constructs to understand the participants' perceptions for privacy vs. convenience based on their rates for privacy and convenience constructs. As such, the Convenience Importance construct quantifies how important the convenience that IoT devices afford to their users. This measure has one direct question for the participant to rate the importance of convenience for them when using smart devices.   

\subsubsection{Privacy Control.}
Using IoT devices generally increases the risks associated with the sensitive personal data transmission, acquisition, and utilization with/ without users knowledge \cite{weinberg2015internet}; therefore, it is critical for users to be able to manage their personal information that IoT devices collected. The Privacy Control construct measured IoT users' ability to control their personal information that are collected by smart devices. This measure has three questions for participants to rate the importance of enabling them to know when and what type of personal information was collected and the importance to require users' permissions prior any data collection by the devices. We used this measure to uncover differences between the participants in terms of their ability to take actions in order to manage the collection of personal information when using smart devices.  

\subsubsection{Privacy Preferences.}
Although IoT have potential benefits, it is also associated with concerns related to recording data of people who does not own the IoT device \cite{jayaraman2017privacy}. The Privacy Preferences measure was developed to quantify users' privacy preferences. This measure has three questions for users to rate the importance of three reactions (not saying anything, hide themselves, or use applications to hide their identity) to protect their privacy in case of a security camera in another place (e.g., friend's house) was recording their audio or video. Thus, we used this measure to examine the participants' differences regarding the importance of their reactions to an IoT device that they do not own, but collected data about them.     

\subsubsection{Convenience Preferences.}
IoT users usually find it difficult to effectively control multiple smart devices \cite{dehury2016design}. Therefore, the Convenience Preferences measure was used to measure users' convenience preferences to manage their smart devices. This measure has three questions for users to rate strategies to manage IoT devices by having one platform, use website, or implement centralized monitoring. Therefore, this measure was important for this study to examine differences on what users would prefer for managing multiple smart devices.  

\subsubsection{Website Satisfaction.}
The Website Satisfaction construct measured participants satisfaction about the proposed website prototype to manage multiple IoT devices. This construct quantifies participants' satisfaction based on: 1) prototype organization, 2) ease of navigation, and 3) user-friendly interface. Measuring users' satisfaction was important to us to evaluate any differences between the participants' regarding their feedback on the website prototype. This measure was also useful to recommend future recommendations that will be discussed in the discussion section.

\subsection {Data Analysis Approach}
To answer RQ1, we first conducted a dependent t-test for paired samples (i.e., within subjects) to examine if there was a significant difference between the importance of privacy and convenience based on participants self-reported scores. We hypothesized the following difference would be detected:

\begin{itemize}
    \item \textbf{H1:} \textit{Participants will rate their Privacy as more important relative to their Convenience.}
\end{itemize}

We then used a mean split to divide the participants into four quadrants based on their scores on the privacy and convenience importance measures. We did this by calculating the mean of the Privacy and Convenience Importance measures, which are listed in Table \ref{stats}. Participants who had higher scores than the mean scores were assigned to the high (privacy or convenience) groups, while  participants who reported scores lower than the mean scores were assigned to the low (privacy or convenience) groups. 

To answer RQ2, we calculated the between group differences on our privacy and convenience preference constructs (i.e., Privacy Control, Privacy Preferences, and Convenience Preferences) based on the four groups. We hypothesized the following between-group differences.

\begin{itemize}
    \item \textbf{H2:} \textit{Participants with low Privacy/ low Convenience Importance will rate a) Privacy Control and b) Privacy Preference lower than participants with high Privacy/ high convenience Importance.}
    \item \textbf{H3:} \textit{Participants with low Privacy/ Convenience Importance will rate a) Convenience Preference lower than participants with high Privacy/ Convenience Importance.}
\end{itemize}

We investigated the differences in the self-reported measures between the generated quadrant groups by conducting ANOVA tests \cite{st1989analysis}. In order to compare individual groups with one another, we conducted post-hoc analyses \cite{hilton2006statnote} for the significant differences found. These identified differences demonstrate a holistic understanding of the distinct patterns between the privacy/convenience groups regarding a series of privacy control and convenience preferences.

\subsection{The Prototype Usability Evaluation}
Based on the tasks described above, we coded whether the participants were able to successfully complete each task or not (0=Incomplete; 1=Complete). In the results section, we present the percentages of the correct and wrong answers regarding these tasks. We also assessed between-group differences on the Website Satisfaction measure based on the Privacy and Convenience Importance groups as a grouping variable (Website Satisfaction). We hypothesized the following significant difference:

\begin{itemize}
    \item \textbf{H4:} \textit{Participants with low privacy/ high convenience will be more satisfied about the website prototype than participants with high privacy / low convenience Importance.}
\end{itemize}

In the next section, we describe how we recruited participants and summarize their demographic profiles.

\subsection{Participant Recruitment and Demographics}
Overall, we recruited 43 participants who completed the study voluntary without a compensation, which was stated in the informed consent prior participating to the study. We recruited participants who are over 18 years old. We advertised through word of mouth, recruitment emails, and by posting the flyers on social media. The study took place online where the participants were given the web URL to interact with the prototype and complete the surveys. Three of the participants answered the survey questions but skipped completing the tasks. Since our informed consent allowed participants to skip questions, we retained their survey data for the RQ1 and RQ2 analyses. Therefore, 38 out of the 43 participants completed the tasks and answered the related questions. 

A diverse sample of participants participated in this study, where 60\%, $N=26$ were between 25-34, 14\%, $N=6$ of them were 18-24, 14\% were 35-44, and 12\% were 45-54 years old. In terms of education, most of our participants completed their bachelor’s degree ($49\%, N=21$) or master’s ($33\%, N=14$) degree programs. Almost all participants ($98\%, N=42$) owned more than two smart home devices. The most frequently chosen device were smartphones with a percentage of 90.70\%, followed by smart watch and smart TV with an equal percentage of 55.81\%.

\begin{table}[hbt!]
\centering
\caption{Participants’ Demographic Information ($N=43$)}
\label{tab:demographic}
\begin{tabular}{lcc}
 \hline
\textbf{Demographic}                                                                         & \textbf{Number} & \textbf{Percentage}                                            \\  \hline
\textbf{Age}                                                                                          &                 &                                                       \\ \hline
18-24                                                                                        & 6               & 13.95\%                                      \\
25-34                                                                                        & 26              & 60.47\%                                      \\
35-44                                                                                        & 6               & 13.95\%                                      \\
45-54                                                                                        & 5               & 11.63\%                                      \\ \hline
\textbf{Educational Background}                                                              &                 &                                                       \\ \hline
High school                                                                         & 6               & 13.95\%                                      \\
Bachelor’s degree                                                                   & 21              & 48.84\%                                      \\
Master’s degree                                                                     & 14              & 32.56\%                                      \\
Ph.D. or higher                                                                     & 2               & 4.65\%                                       \\ \hline
\textbf{smart Devices Owned}                                                                   &                 &                                                       \\ \hline
Smartphone                                                                          & 39              & 90.70\%  \\
Smartwatch                                                                          & 24              & 55.81\%  \\
Activity tracker                                                                    & 10              & 23.26\%  \\
Smart refrigerator                              & 4               & 9.30\%   \\
Smart speaker & 7               & 16.28\%  \\
Smart thermostat                                & 3               & 6.98\%   \\
Smart TV                                        & 24              & 55.81\%  \\
None                                                                                & 1               & 2.33\%  \\  \hline                                  
\end{tabular}
\end{table}

\section{Results}

In Table \ref{stats}, we report the descriptive statistics of our survey measures. As shown in the table, all measures demonstrated adequate construct validity (Cronbach's alpha $>$ 0.70). A general trend we observed was that the means for the privacy-related constructs were typically higher than those associated with convenience-related measures. Further, we note that all constructs were rated relatively high with means ranging from 3.6 to 4.2 on a 5-point scale.

\begin{table}[]
\centering
\caption{The Constructs' Descriptive Statistics}
\label{stats}
\begin{tabular}{lllll}
\hline
\textbf{Constructs}              & \textbf{\begin{tabular}[c]{@{}l@{}}Number \\ of Items\end{tabular}} & \textbf{\begin{tabular}[c]{@{}l@{}}Cronbach's\\ alpha\end{tabular}} & \textbf{Mean} & \textbf{SD} \\ \hline
\textbf{Privacy Importance}      & 1.00                                                                & N/A                                                                 & 4.16          & 0.94        \\
\textbf{Convenience importance}  & 1.00                                                                & N/A                                                                 & 3.88          & 0.82        \\
\textbf{Privacy Control}         & 3.00                                                                & 0.75                                                                & 4.15          & 0.80        \\
\textbf{Privacy Perferences}     & 3.00                                                                & 0.76                                                                & 3.60          & 1.19        \\
\textbf{Convenience Preferences} & 3.00                                                                & 0.80                                                                & 3.85          & 0.86        \\
\textbf{Website Satisfaction}    & 3.00                                                                & 0.83                                                                & 4.20          & 0.79        \\ \hline
\end{tabular}
\end{table}

\subsection{Privacy over Convenience (RQ1)}

A dependent t-test for paired samples yielded a significant difference between users in terms of Privacy Importance and Convenience Importance ($p = 0.01$). Users’ perceived privacy  ($m = 4.23$) as more important than ($p = 0.01$) than their desire for convenience ($m = 3.86$) as shown in table \ref{tab:privacy-convenience}. This results supports our H1.

\begin{table}[hbt!]
\centering
 \caption{Privacy versus Convenience}
\label{tab:privacy-convenience}
\begin{tabular}{lcllc}
  \hline
                                                                     &     & \textbf{Mean} & \textbf{ } & \textbf{SD}                                            \\  \hline
Privacy  &   &         4.23   &  &  0.92  \\ 
 Convenience   &   & 3.86      &         & 0.89   \\ \hline                                  
 \end{tabular}
 \end{table}

The privacy/ convenience quadrant groups resulted in: A) high privacy/ high convenience group, B) high privacy/ low convenience group, C) low privacy/ high convenience group, and D) low privacy/ low convenience group. Table \ref{quad} showed the participants' distribution across the quadrant groups. The largest group among the quadrants was participants who reported both high privacy and high convenience ($N=16, 37\% $) while the smallest group was participants who reported high privacy, but low convenience ($N=5, 12\%$).  
\begin{table}[]
\centering
\caption{The Distribution of Privacy/Convenience Quadrant Groups. The percentages out of the total number of participants ($N=43$). }
\begin{tabular}{llll}
\cline{1-4}
                      & \textbf{High Convenience} & \textbf{Low  Convenience} & \textbf{Total} \\ \cline{1-4} 
\textbf{High Privacy} & N= 16, 37\%                        & N=5, 12\%                         & N=21, 49\%             \\
\textbf{Low Privacy}  & N=12, 28\%                       & N=10, 23\%                        & N=22, 51\%             \\
\textbf{Total}        & N=28, 65\%                        & N=15, 35\%                        &            \\ \hline
\end{tabular}
\label{quad}
\end{table}

To some extent, our results highlight that the trade-off between privacy and convenience may be a false dichotomy, as the majority of our participants felt that both were important. When there was a discernible trade-off, participants tended to prefer convenience over privacy (between-groups), which conflicted with our earlier within-subject findings that individuals tended to rate privacy as more important than convenience when making comparative decisions. 

\subsection{Differences in Privacy Control, Privacy Preferences, and Convenience Preferences (RQ2)}
This section presents the between-group results to examine the differences between the four groups of privacy/convenience importance in terms of Privacy Control, Privacy Preferences, Convenience Preferences, and Website Satisfaction. Table \ref{Means} listed the means and standard deviations of these measures for the four groups. Table \ref{ANOVA} showed significant differences in terms of Privacy Control and Convenience Preferences measures based on ANOVA tests. We will discuss the results of this ANOVA tests in the following subsections.

\begin{table}[hbt!]
\centering
 \caption{Mean and standard deviation of Privacy Control, Privacy Preferences, Convenience Preferences by the privacy/convenience quadrants}
\label{Means}
\begin{tabular}{p{0.25\linewidth} p{0.17\linewidth} p{0.17\linewidth} p{0.18\linewidth} p{0.16\linewidth}}
  \hline
\textbf{Groups}  &  \textbf{Privacy Control} & \textbf{Privacy Preferences } & \textbf{Convenience Preferences } & \textbf{Website Satisfaction} \\  \hline

 & \textbf{M} \hspace{0.2cm} \textbf{SD} & \textbf{M} \hspace{0.2cm} \textbf{SD} & \textbf{M} \hspace{0.2cm} \textbf{SD} & \textbf{M} \hspace{0.2cm} \textbf{SD} \\ \hline
\textbf{High-Privacy/ High-Convenience}  & 4.17 \hspace{0.1cm} 0.44                  & 3.92 \hspace{0.1cm} 0.90 & 4.31 \hspace{0.1cm} 0.74 & 4.25 \hspace{0.1cm} 0.64 \\ 
& & & & \\
\textbf{High-Privacy/ Low-Convenience}  & 4.33 \hspace{0.1cm} 0.67                  & 3.80 \hspace{0.1cm} 1.02                                         & 4.33 \hspace{0.1cm} 0.62 & 4.47 \hspace{0.1cm} 0.38 \\
& & & & \\
\textbf{Low-Privacy/ High-Convenience}  & 3.94 \hspace{0.1cm} 0.68                  & 3.50\hspace{0.1cm} 1.05 & 3.53 \hspace{0.1cm} 0.77 & 4.03 \hspace{0.1cm} 0.70 \\
& & & & \\
\textbf{Low-Privacy/ Low-Convenience}   & 3.43 \hspace{0.1cm} 0.50                  & 3.10 \hspace{0.1cm} 0.86 & 3.27 \hspace{0.1cm} 0.90 & 4.20 \hspace{0.1cm} 0.74 \\ \hline                                 
 \end{tabular}
 \end{table}

\subsubsection{Privacy Control} 
An ANOVA yielded significant differences between the privacy/ convenience groups regarding their Privacy Control ($F(3, 43) = 11.75, p< 0.001$) as shown in Table \ref{ANOVA}. Post-hoc tests (Table \ref{Post}) demonstrated that users in the low privacy/convenience group ($m=3.43$) reported significantly less Privacy Control than the group of high privacy/convenience ($m=4.71$) and the group of high privacy/low convenience ($m=4.33$).Based on this result, the hypothesis \textbf{H2} was supported. We also found that IoT users in the group of low privacy and high convenience ($m=3.43$) had significantly less Privacy Control than the group of high privacy and convenience ($m=94$) as shown in Table \ref{Means}. This partially supported the \textbf{H2} hypothesis since the group of low privacy/ high convenience has only one low privacy group. The group of low privacy/convenience reported less than the average score of the Privacy and Convenience Importance constructs, which align well with their low Privacy Control as well. 

\begin{table}[]
\centering
\caption{ANOVA results for the Privacy Control, Privacy Preferences, Convenience Preferences by the privacy/convenience groups. There were significant differences found between the groups in terms of Privacy Control and Convenience Preferences. Bold values denote significant difference results.}
\begin{tabular}{llll}
\hline
\textbf{Constructs}              & \textbf{F} & \textbf{df} & \textbf{$p-value$} \\ \hline
\textbf{Privacy Control}         &\textbf{ 11.75}      & \textbf{3 }          & \textbf{\textless{}0.001 }\\
\textbf{Privacy Preferences}     & 1.67       & 3           & 0.18             \\
\textbf{Convenience Preferences} & \textbf{5.11}      & \textbf{3 }          & \textbf{0.004}           \\ 
\textbf{Website Satisfaction} & 0.57       & 3           & 0.63           \\ \hline
\end{tabular}
\label{ANOVA}
\end{table}

\begin{table}[]
\caption{Post-hoc tests to identify the significant differences between the privacy/convenience groups.}
\label{Post}
\centering
\begin{tabular}{p{3cm}p{7cm}l}
\hline
\textbf{Constructs}                               & \textbf{Significant Pairwise Differences (Mean)}                                                      & \textbf{p-value} \\ \hline
\textbf{Privacy Control}         & Low Priv./Low Conv.  \textless{}   High Priv./High Conv. & \textless{}0.001 \\
                                                  & Low Priv./Low Conv.  \textless{} High Priv./Low Conv.        & 0.004            \\
                                                  & Low Priv./High Conv.  \textless{} High Priv./High Conv.          & 0.02             \\ \hline
\textbf{Convenience Pref.} & Low Priv./Low Conv.  \textless{} High Priv./High Conv.           & \textless{}0.001 \\
                                                  & Low Priv./High Conv.   \textless{} High Priv./High Conv.            & 0.05             \\ \hline
\end{tabular}

\end{table}

\subsubsection{Privacy Preferences} 
Regarding Privacy Preferences, an ANOVA did not yield any significant differences between the four groups ($F(3, 43) = 1.67, p= 0.18$) as shown in Table \ref{ANOVA}. The mean scores of these groups listed in Table \ref{Means}. This suggests that all groups were fairly high, in the range of ``Neutral'' to ``Somewhat important'' in the Privacy Preferences scale. Therefore, this may be why we did not detect significant differences. Based on this non-significant result, hypothesis \textbf{H2} could not be supported.

\subsubsection{Convenience Preferences}
There were significant differences between the privacy/convenience groups in terms of their Convenience Preferences ($F(3, 43) = 5.11, p= 0.004$) as shown in Table \ref{ANOVA}. Post-hoc analysis showed (Table \ref{Post}) that the group of low privacy/convenience ($m=3.27$) reported significantly less Convenience Preferences than the group of high privacy/convenience ($m=4.31$). Thus, hypothesis \textbf{H3} was supported. Additionally, the group of low privacy/high convenience ($m=3.53$) were significantly lower on the Convenience Preferences scale than the group of high privacy/convenience ($m=4.31$) as shown in Table \ref{Means}. This result partially supported the \textbf{H2} hypothesis. Generally, we found that the participants' self-reported Importance of Privacy and Convenience were fairly aligned with their Convenience Preferences since the group who reported low Privacy and Convenience Importance had the lowest Convenience Preferences. 

 \subsection{Participants’ Evaluation of the Website Usability (RQ3)}
 
 \subsubsection{Website Satisfaction Survey.}
There were no significant differences between the groups based on their website satisfaction ($F(3, 43) =  0.57, p=0.63$) as shown in Table \ref{ANOVA}. Thus, we could not support \textbf{H4} hypothesis because of the non-significant difference. Generally, most participants were satisfied with the website as shown in Table \ref{MeanI}, where all mean scores on the individual items as well as the overall construct (Table \ref{Means}) were higher than 4, which were between ``Very Satisfied'' and ``Satisfied'' on the scale. 

\begin{table}[]
\centering
\caption{The means and standard deviation of the individual scale items for Website Satisfaction.}
\label{MeanI}
\begin{tabular}{lll}
\hline
\textbf{Website Satisfaction Items} & \textbf{Mean} & \textbf{SD} \\ \hline
\textbf{Website organization}       & 4.07          & 0.80        \\
\textbf{Ease of website navigation} & 4.23          & 0.81        \\
\textbf{User friendly interface}    & 4.30          & 0.64        \\ \hline
\end{tabular}
\end{table}

Since we did not find significant differences between the privacy/convenience groups based on Website Satisfaction construct, we went beyond the high satisfaction on the website to evaluate their actual usability of the website prototype, which will be discussed in the next section.

 \subsubsection{Task Completion.}
Next, we present participants' responses to the tasks we assigned to them to evaluate their understanding of the website. In the first task, participants were asked to determine the number of currently connected devices on the website account, 35 out of 38 participants (92.11\%) answered this question correctly. Where the correct answer was six devices.

The second task contained instructions for the participants, where they were asked to turn on the sensor readings function for the temperature and pressure sensor (IoT device). Then, they were asked to determine the number of current readings available for the temperature and pressure sensor. Most participants 36 out of 38 (94.74\%) were able to follow the instructions and answer the question correctly. Where the correct answer was five temperature and pressure readings. Participants were also asked to determine the temperature readings for a specific date and time. All participants ($N= 37, 97.37\%$) answered this question correctly, except for one participant.
\section{Discussion}
In this section, we describe the implications of our findings in relation to prior work and provide
design implications of smart home IoT device management systems.

\subsection{Smart Home IoT Trade-offs between Privacy and Convenience}

Previous works have investigated the factors that may affect people’s opinions about IoT adoption \cite{page_internet_2018}.
In the same direction, we investigated our participants’ views on the importance of privacy and convenience when using smart devices.
While previous works such as \cite {clawson_no_2015,epstein_beyond_2016}, emphasized on the privacy as the reason behind the abandonment of technology from users,  Our findings from the H1 hypothesis test confirms the importance of privacy for IoT devices, showing that privacy is more important than convenience for the smart devices users. This implies that users would avoid using IoT devices due to the compromise of user privacy in the way of collecting sensitive personal data. However, based on our results when examining the privacy/convenience quadrants groups, we found that the largest group among the quadrants was participants who valued both high privacy and high convenience. Thus, our results confirm that the trade-off between privacy and convenience creates a false dichotomy, given that most of our participants valued both privacy and convenience. Existing research on IoT mainly focuses on the importance of privacy for IoT users \cite{page_internet_2018}. Therefore, we urge future research to leverage both users' privacy and convenience in order to understand the perceptions of smart IoT devices users toward their privacy concerns and convenience preferences when using IoT devices.


\subsection{Implications for the Design of Smart Home IoT Device Management Systems}

Our findings demonstrated that in general most participants were satisfied with our prototype (website) since they found it to be well-organized, easy to use, and user-friendly.
This result indicates that the proposed prototype could an easily accessible platform, and it could be used easily by IoT user with different levels of education and without much technical experience (based on our diverse participants demographics).
Having a website to manage IoT devices while preserving users’ privacy is key to IoT devices \cite{boeckl_considerations_2019}. Therefore, we recommend that developer would base their user-centered website designs on our prototype since we showed that the website would be useful in fulfilling IoT requirements in terms of privacy and convenience.

By showing the quadrant groups of privacy and convenience, which demonstrated different levels (i.e, high and low) of privacy and convenience. This suggests that different design solutions should be designed based on these groups' privacy concerns or convenience preferences. This is important because by estimating how much users value their privacy or convenience, IoT developers can predict appropriate features that may become sources of competitive advantage in IoT device management platforms. Therefore, we recommend that IoT designers to take into account that users may not perceive privacy and convenience at the same level. This highlights the importance of creating personalized privacy experiences for IoT users based on either initial survey questions to report their preferences or trained machine learning algorithms that would predict users privacy and convenience preferences similar to smart phone personalized permission management algorithms \cite{wisniewski2020predicting}.




\subsection{Limitations and Future Work}

This section outlines limitations of our study that inform future work in the space of smart home IoT. 
We studied younger adults, where 60\% of our participants were between 25 and 34 years old. Therefore, our results could not be generalized to older adult populations who may have a different privacy versus convenience calculus. Therefore, future studies should further study smart home IoT users' preferences towards privacy versus convenience. In our study,  participants were asked to imagine themselves in a hypothetical situation, where they explored our IoT smart home device management website that was connected to a temperature and pressure sensor that was located in the first-author's home. While it was not feasible for us to test our system in the homes of our actual participants, future studies that leverage existing smart home sensors in participants' homes or install such sensors for the purpose of the study would increase the ecological validity of our results. Finally, our web-based prototype, while functional, had limited capabilities. Future studies that build upon our work could go in more depth in regards to feature design that optimizes users' privacy and convenience when managing their IoT smart home devices.

\section{Conclusion}
A large number of smart home IoT devices demands management and control solutions. Moreover, the growing number of connected devices and their inherent constraints motivate the need for efficient smart home IoT device management that focus on users privacy-preserving. Therefore, we conducted a web-based survey and usability study with 43 participants who use IoT devices frequently to: 1) examine their smart home IoT usage patterns and privacy preferences, and 2) evaluate a web-based prototype for smart home IoT device management. The findings confirmed that privacy is more important for the users than convenience when using smart devices, Moreover, based on our prototype evaluation, we found that all participants were generally satisfied with our website prototype and their actual usability evaluation demonstrated that they understand the functionality of the website. Overall, this study provided a rich picture of privacy and convenience preferences of smart home IoT users when using smart home IoT device management website.

%
%
%
\bibliographystyle{splncs04}
\bibliography{mybibliography}

\begin{thebibliography}{10}
\providecommand{\url}[1]{\texttt{#1}}
\providecommand{\urlprefix}{URL }
\providecommand{\doi}[1]{https://doi.org/#1}

\bibitem{noauthor_living_nodate}
Living in a glass house {\textbar} {Proceedings} of the 13th international
  conference on {Ubiquitous} computing,
  \url{https://dl.acm.org/doi/10.1145/2030112.2030118}

\bibitem{noauthor_smart_2016}
Smart homes, comfort and data capture (Jan 2016),
  \url{https://www.pewresearch.org/internet/2016/01/14/scenario-home-activities-comfort-and-data-capture/}

\bibitem{arabo_privacy_2012}
Arabo, A., Brown, I., El-Moussa, F.: Privacy in the {Age} of {Mobility} and
  {Smart} {Devices} in {Smart} {Homes}. In: 2012 {International} {Conference}
  on {Privacy}, {Security}, {Risk} and {Trust} and 2012 {International}
  {Confernece} on {Social} {Computing}. pp. 819--826 (Sep 2012).
  \doi{10.1109/SocialCom-PASSAT.2012.108}

\bibitem{balliu_securing_2019}
Balliu, M., Bastys, I., Sabelfeld, A.: Securing {IoT} {Apps}. IEEE Security
  Privacy  \textbf{17}(5),  22--29 (Sep 2019). \doi{10.1109/MSEC.2019.2914190},
  conference Name: IEEE Security Privacy

\bibitem{boeckl_considerations_2019}
Boeckl, K., Fagan, M., Fisher, W., Lefkovitz, N., Megas, K.N., Nadeau, E.,
  O'Rourke, D.G., Piccarreta, B., Scarfone, K.: Considerations for managing
  {Internet} of {Things} ({IoT}) cybersecurity and privacy risks. Tech. Rep.
  NIST IR 8228, National Institute of Standards and Technology, Gaithersburg,
  MD (Jun 2019). \doi{10.6028/NIST.IR.8228},
  \url{https://nvlpubs.nist.gov/nistpubs/ir/2019/NIST.IR.8228.pdf}

\bibitem{clawson_no_2015}
Clawson, J., Pater, J., Miller, A., Mynatt, E., Mamykina, L.: No longer
  wearing: investigating the abandonment of personal health-tracking
  technologies on craigslist. pp. 647--658. \doi{10.1145/2750858.2807554}

\bibitem{cronbach1955construct}
Cronbach, L.J., Meehl, P.E.: Construct validity in psychological tests.
  Psychological bulletin  \textbf{52}(4), ~281 (1955)

\bibitem{dehury2016design}
Dehury, C.K., Sahoo, P.K.: Design and implementation of a novel service
  management framework for iot devices in cloud. Journal of Systems and
  Software  \textbf{119},  149--161 (2016)

\bibitem{delicato_platform_2014}
Delicato, F., Pires, P., Avila~Barros, T., Batista, T., Costa, B.: A platform
  for integrating physical devices in the internet of things.
  \doi{10.1109/EUC.2014.42}

\bibitem{epstein_beyond_2016}
Epstein, D.A., Caraway, M., Johnston, C., Ping, A., Fogarty, J., Munson, S.A.:
  Beyond abandonment to next steps: Understanding and designing for life after
  personal informatics tool use. In: Proceedings of the 2016 {CHI} Conference
  on Human Factors in Computing Systems. pp. 1109--1113. {CHI} '16, Association
  for Computing Machinery. \doi{10.1145/2858036.2858045},
  \url{https://doi.org/10.1145/2858036.2858045}

\bibitem{gupta_two-fold_2022}
Gupta, S.D., Kaplan, S., Nygaard, A., Ghanavati, S.: A {Two}-{Fold} {Study}
  to {Investigate} {Users}’ {Perception} of {IoT} {Information}
  {Sensitivity} {Levels} and {Their} {Willingness} to {Share}
  the {Information}. In: Meng, W., Katsikas, S.K. (eds.) Emerging
  {Information} {Security} and {Applications}. pp. 87--107. Communications in
  {Computer} and {Information} {Science}, Springer International Publishing,
  Cham (2022). \doi{10.1007/978-3-030-93956-46}

\bibitem{hilton2006statnote}
Hilton, A., Armstrong, R.A.: Statnote 6: post-hoc anova tests. Microbiologist
  \textbf{2006},  34--36 (2006)

\bibitem{jayaraman2017privacy}
Jayaraman, P.P., Yang, X., Yavari, A., Georgakopoulos, D., Yi, X.: Privacy
  preserving internet of things: From privacy techniques to a blueprint
  architecture and efficient implementation. Future Generation Computer Systems
   \textbf{76},  540--549 (2017)

\bibitem{kroger_personal_2021}
Kröger, J.L., Gellrich, L., Pape, S., Brause, S.R., Ullrich, S.: Personal
  information inference from voice recordings: {User} awareness and privacy
  concerns. Proceedings on Privacy Enhancing Technologies  \textbf{2022}(1),
  6--27 (Dec 2021). \doi{10.2478/popets-2022-0002},
  \url{https://www.sciendo.com/article/10.2478/popets-2022-0002}

\bibitem{kunal_smart_2016}
Kunal, D., Tushar, D., Pooja, U., Vaibhav, Z., Lodha, V.: Smart home automation
  using {IoT}, international journal of advanced research in computer and
  communication engineering.  \textbf{Vol. 5}

\bibitem{lau_alexa_2018}
Lau, J., Zimmerman, B., Schaub, F.: Alexa, {Are} {You} {Listening}? {Privacy}
  {Perceptions}, {Concerns} and {Privacy}-seeking {Behaviors} with {Smart}
  {Speakers}. Proceedings of the ACM on Human-Computer Interaction
  \textbf{2}(CSCW),  102:1--102:31 (Nov 2018). \doi{10.1145/3274371},
  \url{https://doi.org/10.1145/3274371}

\bibitem{martin_smart_nodate}
Martin, C.: Smart {Home} {Technology} {Hits} 69\% {Penetration} in {U}.{S}.,
  \url{https://www.mediapost.com/publications/article/341320/smart-home-technology-hits-69-penetration-in-us.html}

\bibitem{mccreary_contextual_2016}
Mccreary, F., Zafiroglu, A., Patterson, H.: The {Contextual} {Complexity} of
  {Privacy} in {Smart} {Homes} and {Smart} {Buildings}, vol.~9752 (Jul 2016).
  \doi{10.1007/978-3-319-39399-57}, pages: 78

\bibitem{molla_people_2019}
Molla, R.: People say they care about privacy but they continue to buy devices
  that can spy on them (May 2019),
  \url{https://www.vox.com/recode/2019/5/13/18547235/trust-smart-devices-privacy-security}

\bibitem{naeini_privacy_2017}
Naeini, P.E., Bhagavatula, S., Habib, H., Degeling, M., Bauer, L., Cranor,
  L.F., Sadeh, N.: Privacy expectations and preferences in an \{{IoT}\} world.
  pp. 399--412.
  \url{https://www.usenix.org/conference/soups2017/technical-sessions/presentation/naeini}

\bibitem{page_internet_2018}
Page, X., Bahirat, P., Safi, M.I., Knijnenburg, B.P., Wisniewski, P.: The
  internet of what? understanding differences in perceptions and adoption for
  the internet of things  \textbf{2}(4),  183:1--183:22. \doi{10.1145/3287061},
  \url{https://doi.org/10.1145/3287061}

\bibitem{piyare_bluetooth_2011}
Piyare, R., Tazil, M.: Bluetooth based home automation system using cell phone.
  In: 2011 {IEEE} 15th {International} {Symposium} on {Consumer} {Electronics}
  ({ISCE}). pp. 192--195 (Jun 2011). \doi{10.1109/ISCE.2011.5973811}, iSSN:
  2159-1423

\bibitem{saeidi_if_2021}
Saeidi, M., Calvert, M., Au, A.W., Sarma, A., Bobba, R.B.: If {This} {Then}
  {That} : {Exploring} users’ concerns with {IFTTT} applets. Proceedings on
  Privacy Enhancing Technologies  \textbf{2022}(1),  166--186 (Dec 2021).
  \doi{10.2478/popets-2022-0009},
  \url{https://www.sciendo.com/article/10.2478/popets-2022-0009}

\bibitem{shrestha_web_2017}
Shrestha, B., Mali, S., Joseph, A., Singh, K.J.: Web and android based
  automation using {IoT} p.~4

\bibitem{silva_m4dniot-networks_2019}
Silva, J.D.C., Rodrigues, J.J.P.C., Saleem, K., Kozlov, S.A., Rabelo, R.A.L.:
  M4dn.{IoT}-a networks and devices management platform for internet of things
  \textbf{7},  53305--53313. \doi{10.1109/ACCESS.2019.2909436},
  \url{https://ieeexplore.ieee.org/document/8681396/}

\bibitem{st1989analysis}
St, L., Wold, S., et~al.: Analysis of variance (anova). Chemometrics and
  intelligent laboratory systems  \textbf{6}(4),  259--272 (1989)

\bibitem{tabassum_smart_2020}
Tabassum, M., Kropczynski, J., Wisniewski, P., Lipford, H.R.: Smart {Home}
  {Beyond} the {Home}: {A} {Case} for {Community}-{Based} {Access} {Control}.
  In: Proceedings of the 2020 {CHI} {Conference} on {Human} {Factors} in
  {Computing} {Systems}, pp. 1--12. Association for Computing Machinery, New
  York, NY, USA (Apr 2020), \url{https://doi.org/10.1145/3313831.3376255}

\bibitem{weinberg2015internet}
Weinberg, B.D., Milne, G.R., Andonova, Y.G., Hajjat, F.M.: Internet of things:
  Convenience vs. privacy and secrecy. Business Horizons  \textbf{58}(6),
  615--624 (2015)

\bibitem{wisniewski2020predicting}
Wisniewski, P., Safi, M.I., Patil, S., Page, X.: Predicting smartphone
  location-sharing decisions through self-reflection on past privacy behavior.
  Journal of Cybersecurity  \textbf{6}(1),  tyaa014 (2020)

\bibitem{zhang_survey_2014}
Zhang, P., Vasilakos, A.: A {Survey} on {Trust} {Management} for {Internet} of
  {Things}. Journal of Network and Computer Applications  \textbf{42} (Jun
  2014). \doi{10.1016/j.jnca.2014.01.014}

\bibitem{zheng_user_2018}
Zheng, S., Apthorpe, N., Chetty, M., Feamster, N.: User {Perceptions} of
  {Smart} {Home} {IoT} {Privacy}. Proceedings of the ACM on Human-Computer
  Interaction  \textbf{2}(CSCW),  200:1--200:20 (Nov 2018).
  \doi{10.1145/3274469}, \url{https://doi.org/10.1145/3274469}

\end{thebibliography}
%

\newpage

\section{Appendix A}
\setcounter{table}{0} 
\renewcommand{\thetable}{A.\arabic{table}}
\begin{table*}[h]
 \centering
 \footnotesize
 \caption{Survey Items of IoT Device Usage}
   \label{tab:iot-usage}
\begin{tabular}{ |p{12cm}|  }
 \hline
\textit{Which Internet of Things (IoT) device(s) do you own. (Select all that apply)} \\ \hline
1. Smart phone  \\ 
2. Smart watch  \\ 
3. Activity tracker \\ 
4. Smart refrigerator \\ 
5. Smart speaker \\ 
6. Smart thermostat \\ 
7. Smart TV \\ 
8. None\\ 
9. Other (Please specify) \\ \hline
\textit{How many hours per week do you use IoT devices?} \\ \hline
0 hr \\
4-6 hrs \\
7-10 hrs \\ 
11-14 hrs \\
15-20 hrs \\
20+ hrs \\ \hline
\textit{For what purposes do you use IoT devices? (Select all that apply)} \\ \hline 
1. Smart Home \\
2. Smart energy monitoring system \\
3. Vehicle Tracking \\
4. Entertainment \\
5. Lifestyle \\
6. Health monitoring \\
7. None ( do not have an IoT device) \\
8. Other (please specify) \\ \hline
\textit{Which of the following applications do you use to manage your IoT devices?} \\ \hline
1. Wink \\
2. SimpliSafe Home Security \\
3. Yonomi \\
4. ADT Control \\
5. Olisto \\
6. None \\
7. Do not have an IoT device \\
8. Other (please specify) \\ \hline
\end{tabular}
\end{table*}

\renewcommand{\thetable}{A.\arabic{table}}
\begin{table*}[h]
 \centering
 \footnotesize
 \caption{Survey Items of Prototype Satisfaction}
   \label{tab:satisfaction}
\begin{tabular}{ |p{12cm}|  }
 \hline
\textit{Based on your experience in our website http://iotprivacycontrol.com/, how satisfied are you with the following. (1 = Not Satisfied at all, 5 = Very Satisfied)} \\ \hline
1. Website organization  \\ 
2. Ease of website navigation  \\ 
3. User friendly interface \\ \hline
\end{tabular}
\end{table*}

\newpage
\section{Appendix B}
\setcounter{table}{0} 
\renewcommand{\thetable}{B.\arabic{table}}
\begin{table*}[h]
 \centering
 \footnotesize
 \caption{Survey Items of Privacy Concern}
  \label{tab:privacy-concern}
\begin{tabular}{ |p{12cm}|  }
 \hline
\textit{In general, how concerned are you about your privacy in the daily activities as the following? (1 = Not at all concerned, 5 = Very concerned )} \\ \hline
1. People knowing your private and personal information  \\ 
2. Walking in a public place which is full of sensors such as, private security camera, traffic microwave radar sensor, etc.  \\ 
3. To be in the background of photos that are taken by strangers \\
4. To be in the foreground of photos that are taken by strangers \\ \hline
\end{tabular}
\end{table*}

\renewcommand{\thetable}{B.\arabic{table}}
\begin{table*}[h]
 \centering
 \footnotesize
 \caption{Survey Items of Importance of Privacy and Convenience}
  \label{tab:convenience-privacy}
\begin{tabular}{ |p{12cm}|  }
 \hline
\textit{Rate how important privacy (e.g., protecting your personal information) is to you when you are using smart devices. (1 = Not important at all, 5 = Very important)}  \\ \hline
\textit{Rate how important convenience (e.g., completing a task such as, increasing the thermostat temperature) is to you when you are using smart devices. (1 = Not important at all, 5 = Very important)} \\ \hline
\end{tabular}
\end{table*}

\renewcommand{\thetable}{B.\arabic{table}}
\begin{table*}[h]
 \centering
 \footnotesize
 \caption{Survey Items of Privacy Actions for Protecting Personal Information}
  \label{tab:iot-privacy}
\begin{tabular}{ |p{12cm}|  }
 \hline
\textit{If you were using some IoT devices, e.g., Smart Thermostat, Smart TV, and Smart phone, what type of information do you think would be captured by these devices? Select all that apply.} \\ \hline
1. Personal Information (e.g., name, address, bank information, etc.)  \\ 
2. Biometric Information (e.g., Fingerprint, Facial Pattern, Voice, etc.)  \\
3. Location Information 
\\ 
4. Weather Information (e.g., temperature degree) \\
5. Audio recordings \\
6. Video recordings \\
7. Health Information (e.g, medical histories, test and laboratory results, mental health conditions, etc.) \\
8. Other (please specify)
\\ \hline
\textit{How important to you are each of the following actions in terms of protecting your personal information that is captured by IoT devices: (1 = Not important at all, 5 = Very important)}  \\ \hline
1. Enabling you to control what information is being collected about you by IoT devices. \\
2. Informing you when personal information about you is being collected by IoT devices. \\
3. Requesting your permission to collect your information by IoT devices before it is collected. \\ \hline
\textit{Assume you are at your friend’s house and they have a security camera which is recording audio and video that is kept for one week. How important to you are each of the following actions in terms of protecting your personal information that is captured by that IoT device. (1 = Not important at all, 5 = Very important)}  \\ \hline
1. I would be very careful of what I do (e.g, act differently). \\
2. I would be very careful of what I say. \\
3. I would sit in blind spots where I am not captured by the security camera. \\
4. I would use technical methods if applicable (e.g, applications, websites) to hide my identity. \\ \hline
\end{tabular}
\end{table*}

\renewcommand{\thetable}{B.\arabic{table}}
\begin{table*}[h]
 \centering
 \footnotesize
 \caption{Survey Items of Privacy Preference}
  \label{tab:privacy-preference}
\begin{tabular}{ |p{12cm}|  }
 \hline
\textit{Rate the extent to which you agree or disagree with the following actions and statements if you were in this situation: You live in a Smart home that contains different IoT devices and sensors which are: Smart Tv, Smart light, Smart Thermostat, and Smart watch) that capture various types of your information (e.g., your personal information, room temperature degree, your heart rate, your TV watching preferences, etc.), and you want to manage your devices, and reduce the risk of privacy breaching: (1 = Strongly disagree, 5 = Strongly agree )} \\ \hline
1. I am concerned about the privacy of data sensed about me when using IoT devices.  \\ 
2. I prefer to use ONE platform (e.g., website) to manage all my IoT devices.  \\ 
3. I prefer to use website to manage my IoT devices rather than a particular application. \\ 
4. For each device I prefer to use its related application for management purposes. \\ 
5. I prefer to implement centralized monitoring for my IoT devices to manage privacy and security issues. \\ 
6. I prefer to update my IoT devices with regular software updates. \\ \hline
\end{tabular}
\end{table*}

\renewcommand{\thetable}{B.\arabic{table}}
\begin{table*}[h]
 \centering
 \footnotesize
 \caption{Survey Items of Privacy Measures}
  \label{tab:privacy-measures}
\begin{tabular}{ |p{12cm}|  }
 \hline
\textit{Rank the following statements in order of importance from 1 to 5. (1 = Not important at all, 5 = Very important )} \\ \hline
1. Governments should provide new rules and laws to regulate IoT devices to protect our privacy when using them.  \\ 
2. IoT devices' manufacturers need to provide software updates and new features constantly for IoT devices to protect our privacy when using them.  \\ 
3. IoT devices' users need to use platforms (e.g., websites and applications) to manage their IoT devices to protect their privacy. \\ \hline
\end{tabular}
\end{table*}


\end{document}